%% file: skeleton.tex
\documentclass{PoS}

\usepackage{mathtools}
\usepackage[utf8]{inputenc}
\usepackage[titletoc]{appendix}
\usepackage{amsmath}
\usepackage{slashed}
\usepackage{amsfonts}
\usepackage{cite}
\usepackage{lmodern}
\usepackage[T1]{fontenc}

\usepackage{array,multirow,graphicx}
\usepackage{float}
\usepackage{fontenc}
\usepackage{amssymb}
\usepackage{ifthen}

\DeclareOldFontCommand{\rm}{\normalfont\rmfamily}{\mathrm}
\DeclareOldFontCommand{\sf}{\normalfont\sffamily}{\mathsf}
\DeclareOldFontCommand{\tt}{\normalfont\ttfamily}{\mathtt}
\DeclareOldFontCommand{\bf}{\normalfont\bfseries}{\mathbf}
\DeclareOldFontCommand{\it}{\normalfont\itshape}{\mathit}
\DeclareOldFontCommand{\sl}{\normalfont\slshape}{\@nomath\sl}
\DeclareOldFontCommand{\sc}{\normalfont\scshape}{\@nomath\sc}

\def\cpc#1#2#3{{\it Comp.~Phys.~Commun.~}\jref{\bf #1}{#2}{#3}}
\def\epjc#1#2#3{{\it Eur.\ Phys.\ J.\ }\jref{\bf C #1}{#2}{#3}}

\def\jcp#1#2#3{{\it J.~Comp.~Phys.~}\jref{\bf #1}{#2}{#3}}

\def\jhep#1#2#3{{\small\it JHEP~}\jref{\bf #1}{#2}{#3}}

\def\npb#1#2#3{{\it Nucl.~Phys.~}\jref{\bf B #1}{#2}{#3}}

\def\prd#1#2#3{{\it Phys.~Rev.~}\jref{\bf D #1}{#2}{#3}}

\def\prl#1#2#3{{\it Phys.~Rev.~Lett.~}\jref{\bf #1}{#2}{#3}}

\newcommand{\jref}[3]{{\bf #1}, #3 (#2)}
\newcommand{\hepph}[1]{\href{http://arXiv.org/abs/hep-ph/#1}{\texttt{hep-ph/#1}}}

\newcommand{\arxiv}[2]{\href{http://arXiv.org/abs/#1}{\texttt{arXiv:#1\,[#2]}}}
\newcommand{\bibentry}[4]{#1, {\it #2}, #3\ifthenelse{\equal{#4}{}}{}{, }#4.}

\newcounter{notecount}

\title{
\vspace*{-4em}
\mbox{}\hfill \mbox{\small\sc SI-HEP-2019-24, P3H-19-055}\\
\vspace*{4em}
Completing the four-body contributions to $\bar{B}\rightarrow X_s \gamma$ at NLO}

\ShortTitle{Completing the four-body contributions to $\bar{B}\rightarrow X_s \gamma$ at NLO}

\author{Tobias Huber\\
       Naturwissenschaftlich-Technische Fakult\"at, Universit\"at Siegen, \\ Walter-Flex-Str.~3, 57068 Siegen, \\
       E-mail: \email{huber@physik.uni-siegen.de}}

\author{\speaker{Lars-Thorben Moos}\\ 
        Naturwissenschaftlich-Technische Fakult\"at, Universit\"at Siegen, \\ Walter-Flex-Str.~3, 57068 Siegen, Germany\\
        E-mail: \email{moos@physik.uni-siegen.de}}

\abstract{We report on the status of the ongoing calculation of multiparticle contributions to the inclusive radiative $\bar{B} \rightarrow X_s \gamma$ decay at next-to leading order. This effort amounts to the evaluation of the four-particle process $b\rightarrow s \bar{q} q \gamma$ at the one-loop level, supplemented by the corresponding five-particle tree-level cuts $b\rightarrow s \bar{q} q \gamma + g$ of the gluon bremsstrahlung.
Knowledge of these pieces will formally complete the $\bar{B} \rightarrow X_s \gamma$ decay at the next-to-leading order.
The different steps such as the generation of the amplitude, its renormalization, and the treatment of occuring IR-divergences are discussed.
Moreover, we elaborate on the subtleties that arise when treating $\gamma_5$ in $D$ dimensions.}

\FullConference{14th International Symposium on Radiative Corrections (RADCOR2019)\\ 
9-13 September 2019\\
		Palais des Papes, Avignon, France}

\begin{document}

\section{Introduction}
The inclusive radiative decay of the B meson $\bar{B}\rightarrow X_s \gamma$ constitutes one of the most precise tests of the Standard Model (SM) in the quark flavor sector and represents a standard candle in the search for New Physics.\\
At the partonic level, the main contribution comes from the two-particle $b \to s\gamma$ process, which is a flavor-changing neutral current (FCNC) and hence forbidden at tree-level in the SM. Being loop-induced, the process is very sensitive to virtual contributions from new particles running in the loop.

The value for the branching fraction has been measured very precisely. The current experimental value of the CP- and isospin-averaged branching ratio of $\bar{B}\rightarrow X_s \gamma$ with a photon-energy cut of $E_{\gamma} > E_0 = 1.6$~GeV is measured with a precision of
$\sim 5 \%$~\cite{bsgex}
\begin{equation}
\mathcal{B}_{s \gamma}^{exp}=(3.32 \pm 0.15) \cdot 10^{-4} \, .
\end{equation}
With uncertainties on the experimental side that are this small, the results needs to be supplemented accordingly by a theoretical value that is determined with a comparable precision. The work on the theoretical prediction for this process has been carried out for the last twenty years, see e.g.~\cite{burasnlo,misiak06,misiak10,kam12}. This program includes corrections up to next-to-next-to-leading order (NNLO), and resulted in the current SM prediction of the above observable~\cite{bsgth},
\begin{equation}
\mathcal{B}_{s \gamma}^{SM}=(3.36 \pm  0.23) \cdot 10^{-4} \, ,
\end{equation}
which is in very good agreement with the experimental measurement.\\
With the upcoming run of Belle II and the combination with data from the other B-factories, the uncertainties on the experimental side are expected to decrease further, calling for increased effort also on the theory side. In this work, we will focus on the last pieces that are missing in order to formally complete $\bar{B}\rightarrow X_s \gamma$ at the next-to-leading order (NLO). These are multiparticle contributions at the one-loop level, which are suppressed by small CKM factors or Wilson coefficients. To be precise, we elaborate on the one-loop calculation of those four-particle $b\rightarrow s \bar{q} q \gamma$ diagrams that must be supplemented by the corresponding five-particle tree-level cuts $b\rightarrow s \bar{q} q \gamma + g$ from gluon bremsstrahlung. After describing the theoretical framework, the different steps of the computation will be discussed. These include the generation of the diagrams, the Dirac algebra, the reduction of the resulting integrals and their computation, the renormalization and finally the treatment of infrared (IR) divergent collinear pieces that are visible in the final result as logarithms of quark-mass ratios.

\section{Theoretical Framework}
\subsection{Effective Lagrangian}
The interactions that are relevant for the process at hand are incorporated in the following effective Lagrangian,
\begin{equation}
\mathcal{L}_{eff}=\mathcal{L}_{QED+QCD} + \frac{4 G_F}{\sqrt{2}}  \Big[V_{us}^{*}V_{ub}^{\phantom{*}} \sum_{i=1}^2 C_i^u P_i^u+V_{ts}^{*}V_{tb} \sum_{i=3}^6 C_i P_i \Big] \, .
\end{equation}
Here, $\mathcal{L}_{QED+QCD}$ is the QED and QCD Standard Model Lagrangian, the $V_{ij}$ are the entries of the CKM-matrix and the $P_i$ are effective four-fermion operators with Wilson coefficients $C_i$. The operators are given by
\begin{equation}
\begin{alignedat}{4}
P_1^u &= (\bar{s}_L \gamma_{\mu} T^a u_L)(\bar{u}_L \gamma^{\mu} T^a b_L)  &&
P_2^u &&= (\bar{s}_L \gamma_{\mu} u_L)(\bar{u}_L \gamma^{\mu} b_L) \label{eq:op1} \\
P_3 &= (\bar{s}_L \gamma_{\mu} b_L)\sum_{q}(\bar{q} \gamma^{\mu} q)  && 
P_4 &&= (\bar{s}_L \gamma_{\mu} T^a b_L)\sum_{q}(\bar{q} \gamma^{\mu} T^a q)\\
P_5 &= (\bar{s}_L \gamma_{\mu}\gamma_{\nu}\gamma_{\rho} b_L)\sum_{q}(\bar{q} \gamma^{\mu}\gamma^{\nu}\gamma^{\rho} q) \quad && 
P_6 &&= (\bar{s}_L \gamma_{\mu}\gamma_{\nu}\gamma_{\rho} T^a b_L)\sum_{q}(\bar{q} \gamma^{\mu}\gamma^{\nu}\gamma^{\rho} T^a q) \, . \\
\end{alignedat}
\end{equation}
The sum over $q$ runs over five flavours in principle. However, since by definition the $X_s$ system does not contain any charm- or anti-charm quarks and bottom is forbidden kinematically, we can resrict the sum to run over $q=u,d,s$, which we treat as massless. 

\subsection{Contributing Diagrams}

\begin{figure}[t]
\includegraphics[width=0.5\textwidth]{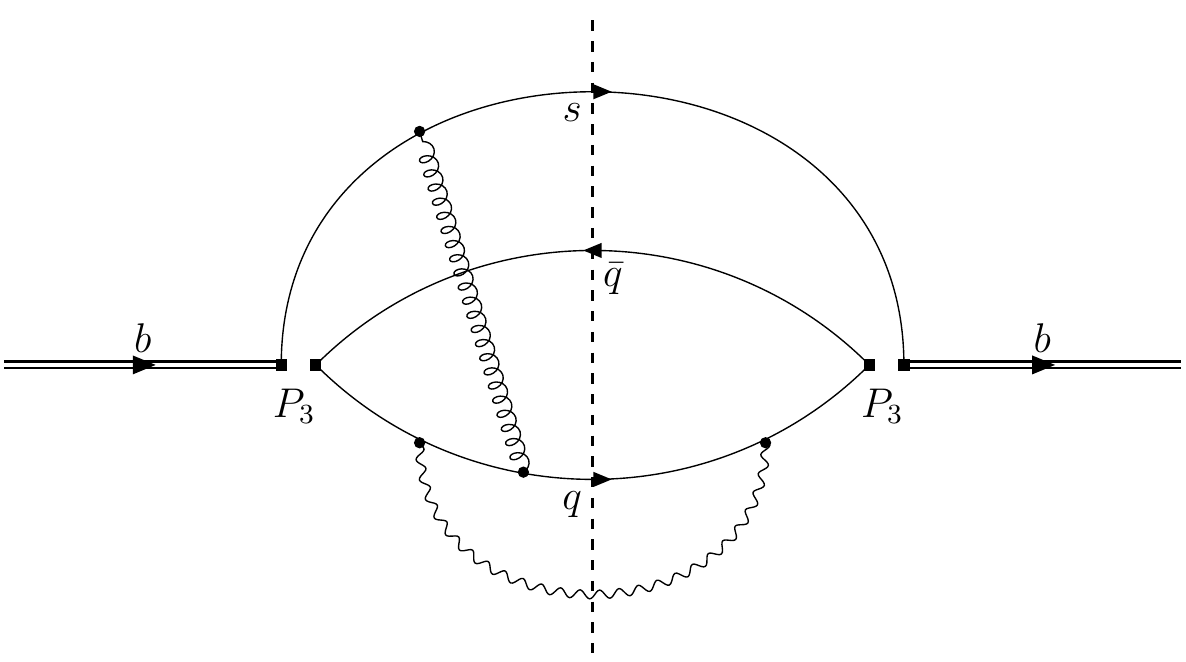}
\includegraphics[width=0.5\textwidth]{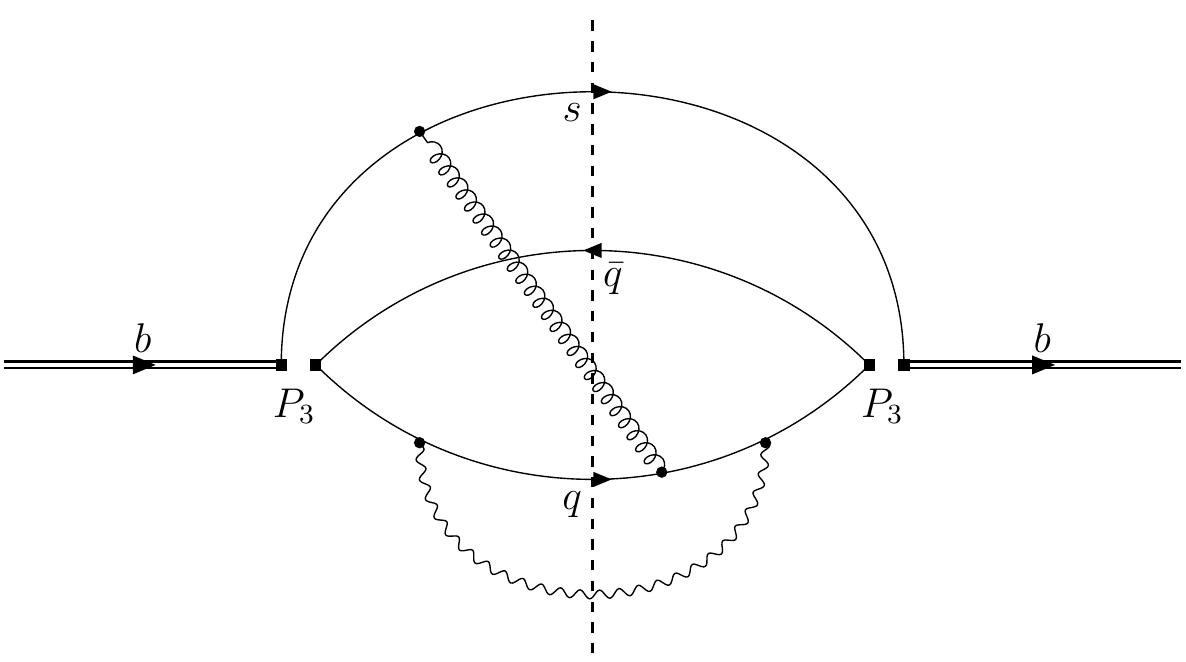}
\begin{center}
\includegraphics[width=0.5\textwidth]{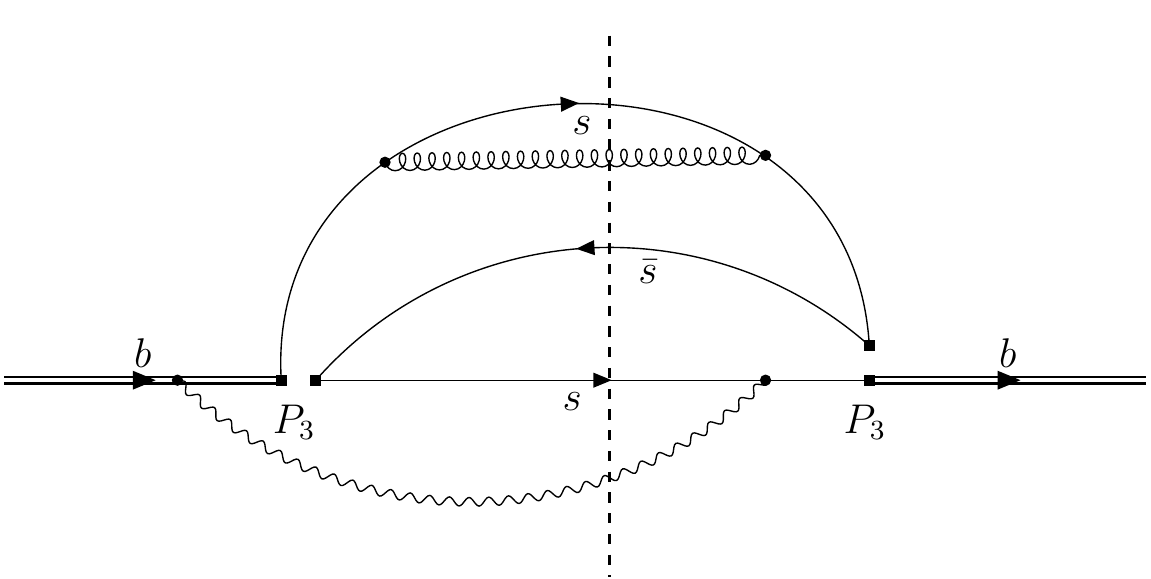}
\end{center}
\caption{Sample diagrams of four-body contributions to the process $ \bar{B} \rightarrow X_s+\gamma$ (top left) and their five-body counterparts (top right). The lower diagram shows the case of a single trace, that occurs when operators with three strange quarks are inserted.}
\label{fig:conts}
\end{figure}

In a previous calculation~\cite{huber:1411}, part of the NLO four-body contribution has already been calculated, namely those pieces that do not require the inclusion of gluon bremsstrahlung. The pieces that remain can be seen in Fig.~\ref{fig:conts}.
For these diagrams, the four-body corrections including the gluon-loop (top left panel) need to be supplemented by the tree-level five-body cuts depicted in the top right panel in order to cancel the IR-divergences in the final state that are induced by the gluon. The lower panel shows an additional operator insertion that appears in the $b\rightarrow s \bar{s} s \gamma (+g)$ channel. Note that this procedure will not cancel all IR divergences, since there are additional IR divergent pieces that remain because of the photon in the final state. The treatment of these will be discussed in a later section.

In our setup, the diagrams are generated with \texttt{QGRAF}~\cite{qgraf} and their computation is carried out in \texttt{FORM}~\cite{form}. After the Dirac algebra and the calculation of the traces, we use Passarino-Veltman decomposition to simplify the results, which is carried out in \texttt{FeynCalc} \cite{feyncalc}. The next step is the further reduction of the result via IBP relations in \texttt{FIRE} \cite{fire} and then finally the phase-space integration of the master integrals in \texttt{Mathematica} and the expansion of the resulting functions in \texttt{HypExp} \cite{hypexp}.

\subsection{Fierz identities for the current-current operators and the treatment of $\gamma_5$}
If we use the operators of Eq. (\ref{eq:op1}) in their original form, we encounter in the squared matrix element products of two traces containing up to two $\gamma_5$ each. In this case, an unambiguous treatment in $D$ dimensions is complicated. We avoid these problems by using Fierz transformations on the operators $P_1^u$ and $P_2^u$~\cite{burasfierz05,huber:1411},
\begin{equation}
P_1^u = - \frac{4}{27} P_3^u + \frac{1}{9} P_4^u + \frac{1}{27} P_5^u - \frac{1}{36} P_6^u + \mathcal{O}(\epsilon) \, ,
\end{equation}
with the notation $P_3^u = (\bar{s}_L \gamma_{\mu} b_L)(\bar{u} \gamma^{\mu} u)$ etc. These relations trade the occurence of an additional projector $P_L$ in the current-current operators for a linear combination of physical penguin operators plus evanescent operators stemming from the fierzing of the fermion lines. The necessary evanescent structure has for example been calculated in Ref.~\cite{burasfierz05} and looks as follows,
\begin{equation}\label{eq:fierz}
 E_1 = (\bar{s}_L \gamma_{\mu}\gamma_{\nu}\gamma_{\rho} T^a u_L)(\bar{u}_L \gamma^{\mu}\gamma^{\nu}\gamma^{\rho} T^a b_L)-(16+4\epsilon)P_1^u \, .
\end{equation}
The corresponding operator for $P_2^u$ only differes by a color factor.

After the Fierz transformation of the operators $P_1^u$ and $P_2^u$, at most a single trace contains zero, one or two occurrences of $\gamma_5$. Despite the fact that $\gamma_5$ is only well-defined in four dimensions we can nevertheless apply the scheme of Naive Dimensional Regularization (NDR) to treat $\gamma_5$ consistently in $D$ dimensions. To this end we use the relation
\begin{equation}
\mathrm{Tr}(\gamma_{\mu_1...\mu_m} \gamma_5 \gamma_{\mu_{m+1}...\mu_n}\gamma_5) = (-1)^{n-m} \mathrm{Tr}(\gamma_{\mu_1...\mu_m \mu_{m+1}...\mu_n})
\end{equation}
by using $\{\gamma_5,\gamma_{\mu}\}=0$ in traces with an {\emph{even}} number of them, together with ($\gamma_5)^2=1$. Traces that do not contain any $\gamma_5$ are then evaluated as products of metric tensors in the usual way.
For the traces with only a {\emph{single}} $\gamma_5$ the case is not as simple. However, since we are computing a squared matrix element, the final result will not have any open Lorentz indices. Using this feature and the cyclicity of the trace (but no anticommutation of $\gamma_5$ in this case!) we can infer that in any term of the squared matrix element $\gamma_5$ appears at most in a single place in a term which has the structure
\begin{equation}
 \mathrm{Tr}( \slashed{p_1}\slashed{p_2}\slashed{p_3}\slashed{p_4}\gamma_5) \, , \label{eq:4gamgam5}
\end{equation}
where the $p_i$ are the light-like momenta of the final state particles.
Traces with fewer than four $\gamma_{\mu}$ and a $\gamma_5$ are consistently set to zero.
Subsequently, we integrate over the phase space whose measure, even in the presence of a photon-energy cut (see below), is sufficiently symmetric to make the antisymmetric structure~(\ref{eq:4gamgam5}) vanish.

\section{Reduction via IBP relations}

\begin{figure}[t]
\hspace{0.15\textwidth}
\includegraphics[width=0.3\textwidth]{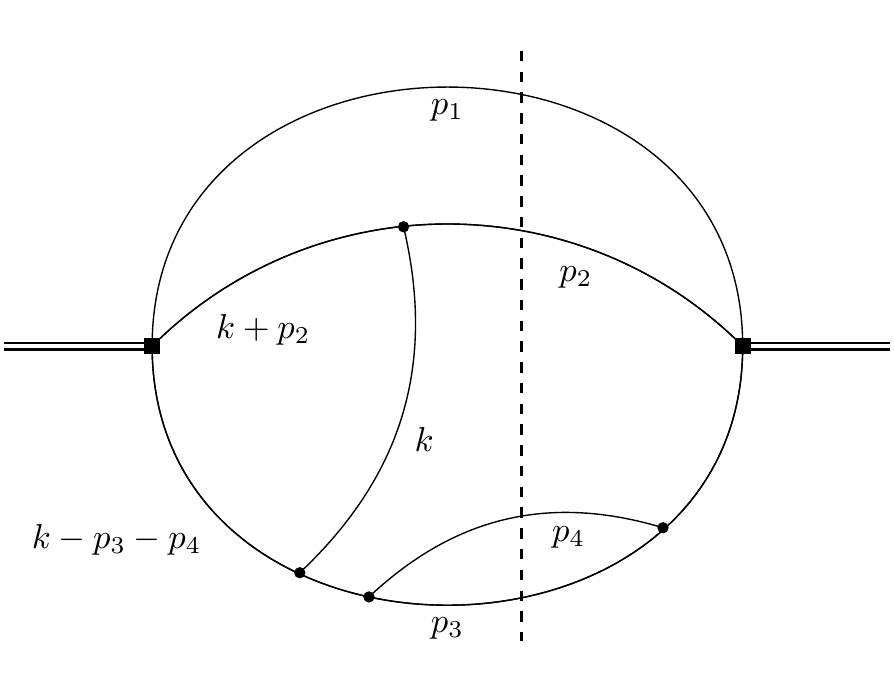}
\hspace{0.1\textwidth}
\includegraphics[width=0.3\textwidth]{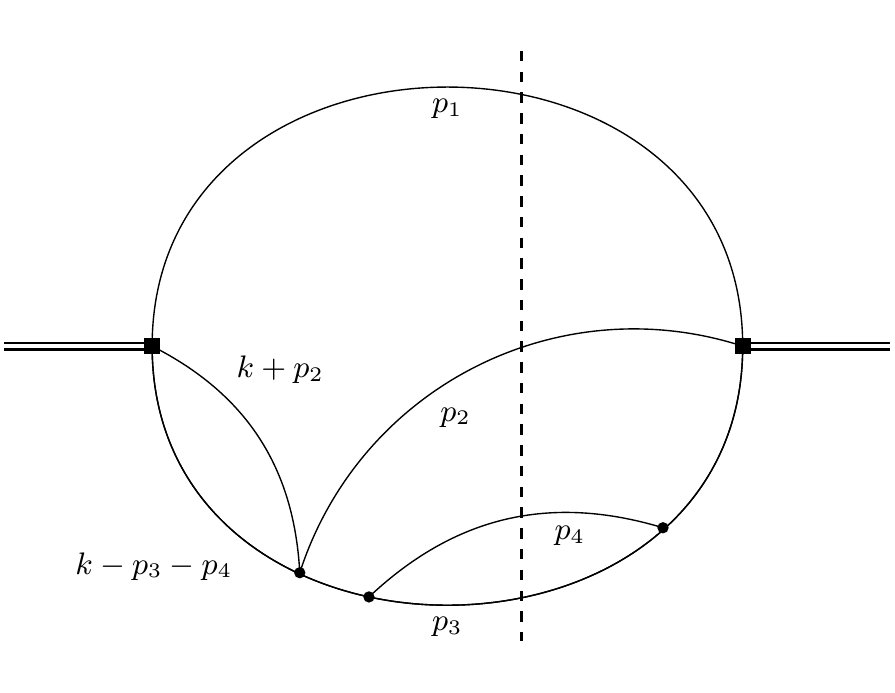}
\caption{Four-loop topologies that occur before (left) and after (right) the reduction. If any propagator
with momentum $p_i$ gets contracted, the corresponding diagram can be set to zero.}
\label{fig:ibp}
\end{figure}

To reduce the number of integrals that need to be solved in the end, we employ an integration-by-parts (IBP) reduction. In order to convert the phase space into a loop integral we employ the relation \cite{melnikovibp}
\begin{equation}
-2\pi i \delta(p^2) = \frac{1}{p^2+i\varepsilon}-\frac{1}{p^2-i \varepsilon} \, , \label{eq:deltaprop}
\end{equation}
which results in four-loop propagator diagrams like the one on the left in Fig.~\ref{fig:ibp}.
The relation~(\ref{eq:deltaprop}) is not only used for the on-shell-conditions $p_i^2=0$ of the outgoing particles, but also
for imposing the condition for the photon-energy cut, of which we give more details in the next section.
The reduction then does not care about the sign of the $i\varepsilon$-prescription and we can run it with any one of the above propagators.
After the reduction procedure, we substitute the occurring propagators back to the $\delta$-functions. It then becomes clear that in
case a propagator with momentum $p_i$ gets contracted, the corresponding integral can be immediately set to zero due the relation $\delta(p^2)p^2=0$.

\section{Phase-space integration}

\subsection{Phase-space measure}

After the computation of the diagrams, we integrate the kernels $\mathcal{K}(s_{ij})=|\mathcal{M}(s_{ij})|^2$ over the four- respectively five-particle phase space in $D=4-2\epsilon$ dimensions. For this we introduce the momentum invariants $s_{ij}$ that are defined by $s_{ij}=2 p_i \cdot p_j / m_b^2$, with the momenta labelled by $ b (p_b) \rightarrow  q( p_1 ) \bar{q}( p_2 ) s( p_3 ) \gamma(p_4) g(p_5).$\\
After the change of variables, the phase-space integral for the four-particle cuts looks as follows~\cite{ps4},
\begin{equation}
\int  [ds_{ij}] \  \delta (1- \sum s_{ij})\mathcal{K}(s_{ij})(-\Delta_4)^{\frac{D-5}{2}}\Theta(-\Delta_4) \, .
\end{equation}
In the formula above, the $\delta (1-\sum s_{ij}) $ incorporates the momentum conservation, while
$\Delta_4$ represents the Gram determinant $\Delta_4= \lambda(s_{12}s_{34},s_{13}s_{24},s_{14}s_{23})$ with
the K\"allen function $\lambda(x,y,z) = x^2 + y^2 + z^2 - 2 x y - 2 x z - 2 y z$. The Gram determinant
can be parametrized according to
\begin{equation}
-\Delta_4=(\bar{z}-s_{34})^2 (a^+ - s_{23})(s_{23} - a^-) \, ,
\end{equation}
where the roots $a^{\pm}$ are given by
\begin{equation}
a^{\pm} = z \big[\bar{v}wx+\bar{x}\bar{w}\pm 2(\bar{v}w\bar{w}x\bar{x})^{1/2} \big] \, .
\end{equation}
Here $z$ is the parameter related to the photon energy which is, alike the other variables, a function of the $s_{ij}$, see below.
Moreover, we use $\bar{z}=1-z$ and similar for the other variables.
A parametrization of the five-particle phase space in $D$ dimensions, including transformations that lead to a factorization of the Gram determinant, are given in~\cite{ps5}.



\subsection{Cut on the photon energy $E_{\gamma}$} \label{section:cut}
Since the measurement of the energy $E_{\gamma}$ of the photon poses a problem in the lower part of the spectrum, we impose a cut on the photon energy to make the prediction compatible with experimental results. In the restframe of the $b$ quark, we have for the photon energy
\begin{equation}
2 E_{\gamma}/m_b = 2 p_b\cdot p_4/m_b^2 = s_{14}+s_{24}+s_{34} \equiv 1-z \, ,
\end{equation}
and the inequality that incorporates the energy cut is expressed as $E_{\gamma}>E_0\equiv\frac{m_b}{2}(1-\delta)$, leading to the relation
\begin{equation}
1-z = s_{14}+s_{24}+s_{34} > 1- \delta.
\end{equation}
To take all this into account in the phase-space integral, the delta function $\delta(1-z-s_{14}-s_{24}-s_{34})$ is added, together with an additional integration over $z$, running from $0$ to $\delta$,
\begin{equation} \label{eq:psint}
\int_{0}^{\delta} dz\int_0^1 [ds_{ij}] \delta (1-z-s_{14}-s_{24}-s_{34}) \delta (z-s_{12}-s_{23}-s_{13}) \ \mathcal{K}(s_{ij}) (-\Delta_4)^{\frac{D-5}{2}} \Theta(-\Delta_4) \, .
\end{equation}

\subsection{Sample kernel}
To illustrate the procedure, we will now sketch the computation of the resulting phase space integrals in the four-particle case. Going through the steps mentioned in the previous sections, one arrives at expressions such as
\begin{equation}
\tilde{I}=\int dPS_4 \int \frac{d^D \ell}{(4 \pi)^D} \frac{s_{13}s_{24}}{\ell^2 (\ell+k_1+k_2+k_3)^2 s_{34}}(-{\Delta_4})^{\frac{D-5}{2}}\Theta(-\Delta_4) \, .
\end{equation}
After the loop integration this evaluates to
\begin{equation}I = \int dPS_4  \frac{\Gamma(\epsilon)\Gamma(1-\epsilon)^2}{\Gamma(2-2\epsilon)} \frac{s_{13}s_{24}(s_{23}+s_{34}+s_{24})^{-\epsilon} }{s_{34}}(-{\Delta_4})^{\frac{D-5}{2}}\Theta(-\Delta_4) \ .
\end{equation} 
To make the subsequent steps easier, one can use the symmetry of~(\ref{eq:psint}) in the momenta of the light quarks (here $1 \rightarrow 3 \rightarrow 2 \rightarrow 1$) prior to conducting the change of variables
\begin{equation}\label{eq:cov}
\begin{alignedat}{4}
& s_{13}  &&= z-s_{23}-s_{12} \, ,\qquad && s_{24}&&= \bar{z}-s_{14}-s_{34}\, , \\
&s_{12} &&= v w z \, , &&s_{34} &&= \bar{z}\bar{v}\, ,\\
&s_{14} &&= \bar{z}v x\, ,  && s_{23}&&=(a^+ - a^-)u + a^-  \, .
\end{alignedat}
\end{equation}
In the above equation, the first two kinematic invariants are fixed by the $\delta$-functions from Eq. (\ref{eq:psint}) and the rest is chosen such that the Gram determinant factorizes. This substitution leads to  the integral
\begin{align}
I & = \int\limits_0^{\delta} dz (z\bar{z})^{D-3} \int\limits_0^1 du\ dv\ dx\ dw\ (u\bar{u})^{\frac{D-5}{2}} v^{D-3} (\bar{v} w \bar{w} x \bar{x})^{\frac{D-4}{2}} \nonumber \\[0.3em]
& \times  \Big[(a^+ -a^-)u+a^-\Big]x \bar{x}^{-1} \Big[v(wz+\bar{z})\Big]^{-\epsilon} \, .
\end{align}
\\
The evaluation of $I$ leads to a sum of hypergeometric functions,
\begin{equation}
I=\int\limits_0^{\delta} dz \ c_1(\epsilon)\ \bar{z}^{2-4\epsilon}z^{2-2\epsilon}\ {}_{2}{\mathrm{F}}_1(\mbox{\small $ 3-3\epsilon,1-\epsilon;3-2\epsilon;z$})+
c_2(\epsilon)\ \bar{z}^{3-4\epsilon}z^{2-2\epsilon}\ {}_{2}{\mathrm{F}}_1(\mbox{\small $3-3\epsilon,2-\epsilon;3-2\epsilon;z$}) \, ,
\end{equation}
where the $c_i$ are functions of the dimensional regulator $\epsilon$. Note that the above expressions are still differential in the photon energy since the integral over $z$ has not yet been carried out.
In the case when a fully analytic expressions to all orders in $\epsilon$ can be achieved, e.g.\ in terms of hypergeometric and $\Gamma$-functions, the integration over $z$ and the expansion in $\epsilon$ can be interchanged, provided the $z$-integration does not lead to further poles in $\epsilon$. The final result can then be obtained as a function of $\delta$ to the desired order in $\epsilon$.
In cases where an all-order result is not possible, one can derive Mellin-Barnes representations, which can be analytically continued to $\epsilon=0$. After expanding in $\epsilon$ and carrying out all remaining integrations, analytic results as functions of $\delta$ can also be achieved in this case.

\section{UV renormalization}

\begin{figure}[t]
\includegraphics[width=0.5\textwidth]{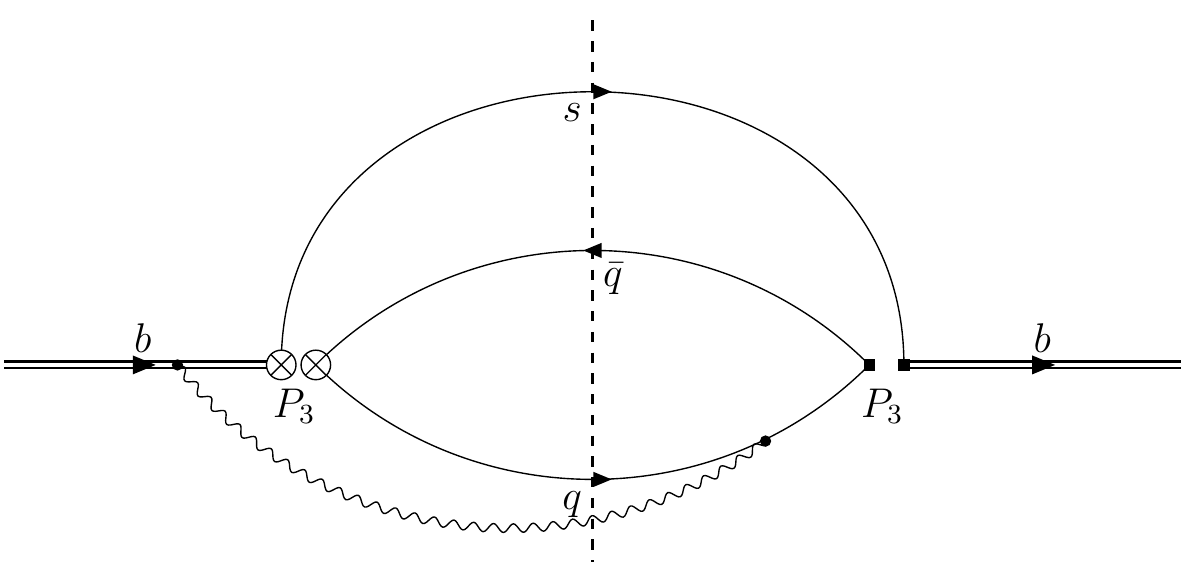}
\includegraphics[width=0.5\textwidth]{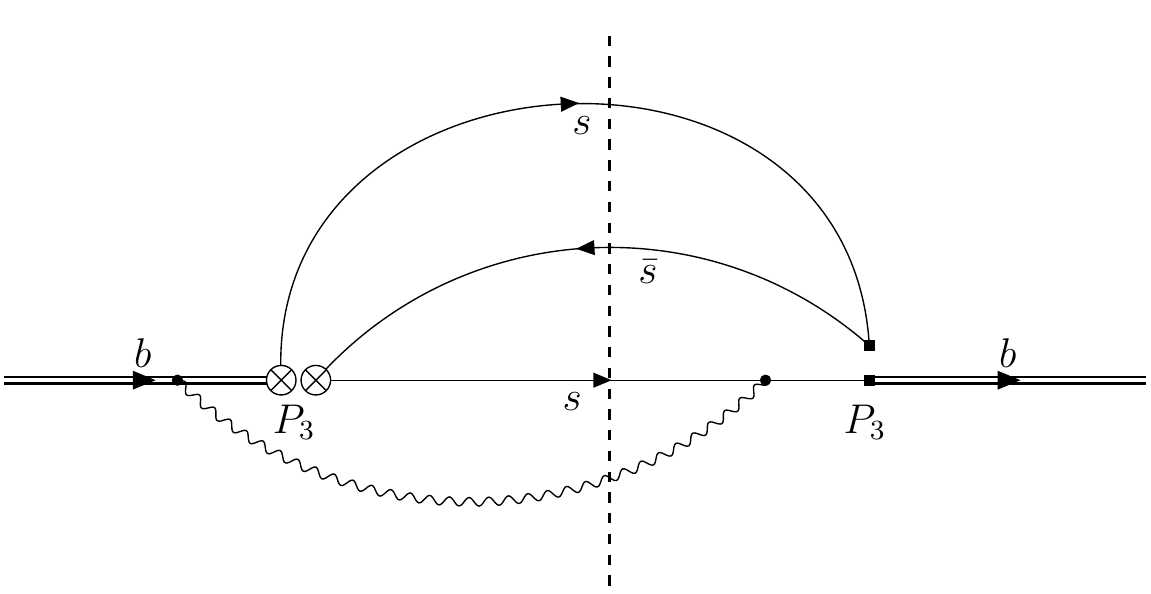}
\caption{Tree-level counterterm diagrams. Insertions of evanescent operators must also be included.}
\label{fig:ct}
\end{figure}

After the phase-space integration, the result has to be renormalized. For this the counterterm insertions in the corresponding four-particle cut diagrams have to be included, see Fig~\ref{fig:ct}. In this step also evanescent operators become important since they can lead to finite pieces in the final result by multiplying $1/\epsilon$-poles from renormalization constants.

\section{Treatment of the collinear IR-divergences}
In regions where the photon is collinear to one of the quarks, we run into collinear divergences. In the massive case these divergences are regulated naturally by the quark masses. Since we work in the case where all the outgoing quarks are massless, these IR-divergences are regularized dimensionally and show up as poles in $\epsilon$. In our case we relate the collinear $1/\epsilon$ poles to logarithms of quark masses by employing the splitting functions. These can be used because in the quasi-collinear limit the amplitude factorizes,

\begin{equation}
 b \rightarrow q_1 q_2 \bar{q}_3 \gamma \Rightarrow b \rightarrow \sum_i q_1 q_2 \bar{q}_3 \times f_i \, .
\end{equation}
In this framework, the $f_i$ is a DGLAP splitting function describing the emission of a photon $\gamma$ from the quark-line $q_i$.

\begin{figure}[t]
\includegraphics[width=0.5\textwidth]{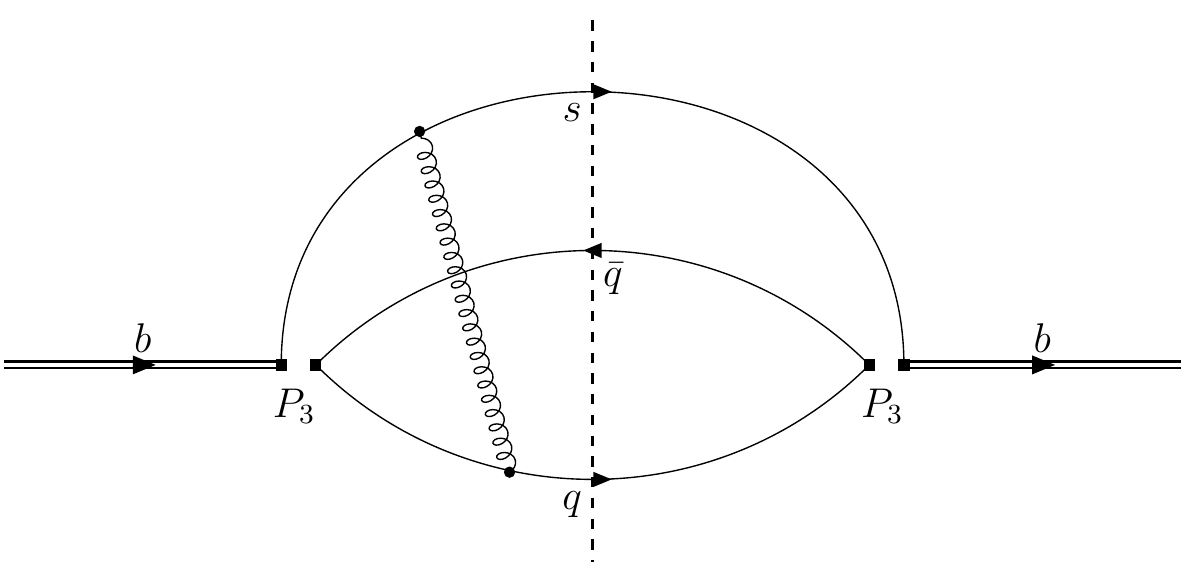}
\includegraphics[width=0.5\textwidth]{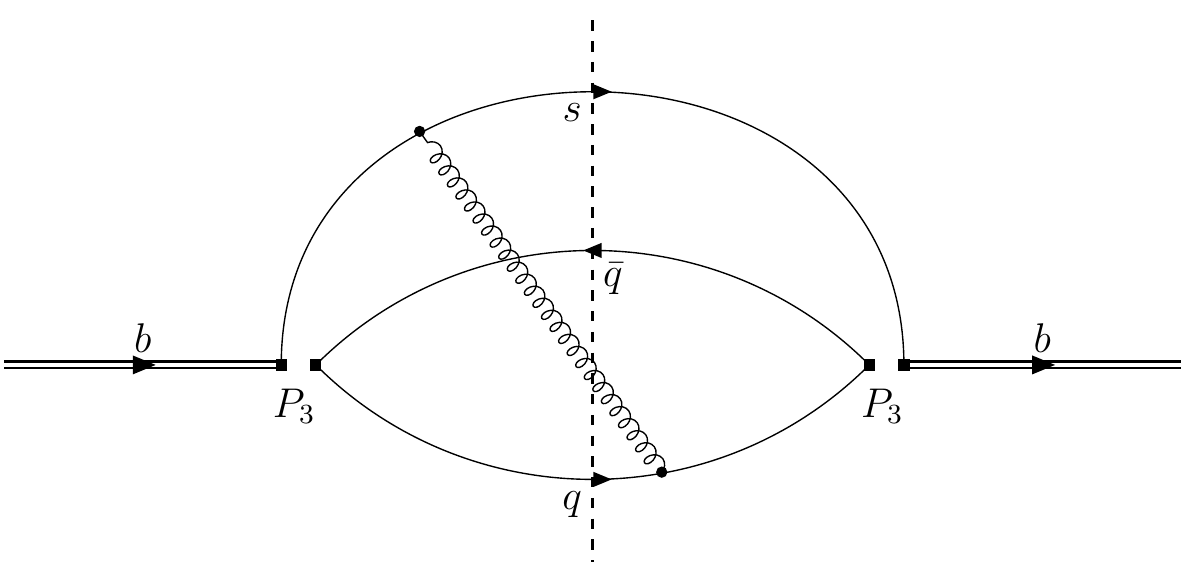}
\begin{center}
\includegraphics[width=0.5\textwidth]{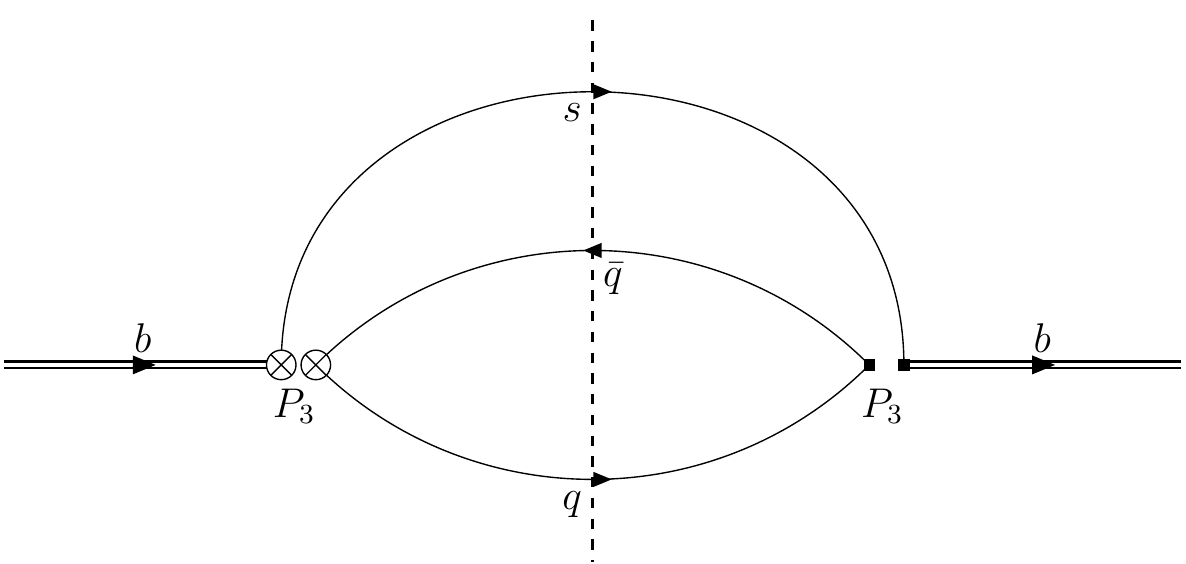}
\end{center}
\caption{Upper two panels: Three and four-particle cut diagrams contributing to the computation of the shift induced by the splitting function. The lower panel shows the relevant counterterm insertions.}
\label{fig:coll}
\end{figure}

A comparison of the splitting functions in the two different schemes yields a shift relation. This relation can then be used to switch from dimensional regularization to the scheme of mass regularization and vice versa,

\begin{equation}
 \frac{d \Gamma_m}{dz} = \frac{d\Gamma_{\epsilon}}{dz}+\frac{d \Gamma_{shift}}{dz} \, .
\end{equation}
The shift part has contributions from three- and four-particle cut diagrams. The shift induced by the three-particle cut diagrams can be computed by means of the following formula,

\begin{equation}
\begin{aligned}
\frac{\Gamma_{shift}}{dz}&=\frac{1}{2 m_b}\frac{1}{2 N_c} \int dPS_3 \mathcal{K}_3(s_{ij})\frac{\alpha_e}{2 \pi \bar{z}} \Bigg\{ Q_1^2 \Bigg[ 1+ \frac{(z-s_{23})^2}{(1-s_{23})^2} \Bigg]  \\ & \quad \times \Bigg[ \frac{1}{\epsilon} -1 +2 \log\frac{(1-s_{23})\mu}{m_{q_1}(1-z)}  \Bigg]\Theta(z-s_{23})  + (\mathrm{cyclic}) \Bigg\} \, .
\end{aligned}
\end{equation}
Sample three- and four-particle cut diagrams and the necessary counterterms can be found in Fig.~\ref{fig:coll}.
Through this shift, we trade the $1/\epsilon$ terms coming from IR divergences for $\log(m_q/m_b)$ terms. In these logarithms, the $m_q$ is not the physical mass of the quarks, but can be varied in a typical range of $\mathcal{O}(100~\mathrm{MeV})$ to get an estimate of the size of the collinear logarithms. 

\section*{Acknowledgments}

We would like to thank the organisers of ``RADCOR 2019'' for creating a very pleasant and inspiring atmosphere, and Miko{\l}aj Misiak for useful correspondence. This research was supported by the Deutsche Forschungsgemeinschaft (DFG, German Research Foundation) under grant  396021762 - TRR 257 ``Particle Physics Phenomenology after the Higgs Discovery''.

\input{bibfinal.tex}

\end{document}

%% file: bibfinal.tex
\newpage{\pagestyle{empty}\cleardoublepage}